\documentclass{article}

\usepackage[utf8]{inputenc}

\usepackage[T1]{fontenc}    
\usepackage{hyperref}       
\usepackage{url}            
\usepackage{booktabs}       
\usepackage{amsmath, amssymb}
\usepackage{amsfonts}       
\usepackage{nicefrac}       
\usepackage{microtype}      
\usepackage{xcolor}         
\usepackage{graphicx}
\graphicspath{ {./figures/} }
\usepackage{subcaption}
\usepackage{cleveref}
\usepackage[colorinlistoftodos,prependcaption,textsize=tiny]{todonotes}
\usepackage[toc=false]{glossaries}
\usepackage{chemformula}
\usepackage{listings}
\usepackage{courier} 
\usepackage{xargs}
\usepackage{tikz}
\usepackage{xspace}

\pdfoutput=1
\usepackage[margin=1in]{geometry}
\usepackage{microtype}
\usepackage[T1]{fontenc}

\newacronym{LLM}{LLM}{Large Language Model}
\newacronym{LSC}{LSC}{Lunar Sample Compendium}
\newacronym{RAG}{RAG}{Retrieval Augmented Generation}
\newacronym{AI}{AI}{Artificial Intelligence}
\newacronym{ppm}{ppm}{parts per million}
\newacronym{ppb}{ppb}{parts per billion}
\newacronym{CSV}{CSV}{Comma Separated Value}
\newacronym{IQR}{IQR}{Interquartile Range}

\crefname{lstlisting}{listing}{listings} 
\Crefname{lstlisting}{Listing}{Listings}

\newcommand{\ChatGPT}{ChatGPT4o\xspace}

\newcommandx{\todomike}[2][1=]{\todo[linecolor=red,backgroundcolor=red!25,bordercolor=red,#1]{#2}}
\newcommandx{\commentmike}[2][1=]{\todo[linecolor=blue,backgroundcolor=blue!25,bordercolor=blue,#1]{#2}}

\title{Towards Large Language Models for Lunar Mission Planning and In Situ Resource Utilization}

\author
{
\and Michael Pekala
\and Gregory Canal
\and Samuel Barham
\and Milena B. Graziano
\and Morgan Trexler
\and Leslie Hamilton
\and Elizabeth Reilly
\and Christopher D. Stiles 
\and \vspace{-2mm} \\ 
 Johns Hopkins University Applied Physics Laboratory \\ 
 11100 Johns Hopkins Road, Laurel, Maryland 20723, USA \\
\texttt{\emph{firstname.lastname}@jhuapl.edu}
}
\date{}

\begin{document}

\maketitle

\begin{abstract}
A key factor for lunar mission planning is the ability to assess the local availability of raw materials.
However, many potentially relevant measurements are scattered across a variety of scientific publications.
In this paper we consider the viability of obtaining lunar composition data by leveraging \glspl*{LLM} to rapidly process a corpus of scientific publications.
While leveraging \glspl*{LLM} to obtain knowledge from scientific documents is not new, this particular application presents interesting challenges due to the heterogeneity of lunar samples and the nuances involved in their characterization.
Accuracy and uncertainty quantification are particularly crucial since many materials properties can be sensitive to small variations in composition. 
Our findings indicate that off-the-shelf \glspl*{LLM} are generally effective at extracting data from tables commonly found in these documents. 
However, there remains opportunity to further refine the data we extract in this initial approach; in particular, to capture fine-grained mineralogy information and to improve performance on more subtle/complex pieces of information.
\end{abstract}

\section{Introduction}

Imagine if we could ask our computers for an overview of natural resources and their characteristics at a specific lunar location. This would be incredibly useful for mission planning and lunar surface infrastructure builds, such as landing pads, habitats, and berms. 
Our long-term vision is \gls*{AI}-powered tools that will support interactive inquiries such as: ``I’m planning a mission to the Moon at this location, and when I arrive, I need to build a habitat. What materials are available within a 2-mile radius of my landing site and in what quantities? How should I process these materials to fabricate a habitat?'' 

\gls*{LLM}-based systems to enable terrestrial travel planning are already being considered by the research community, e.g.,~\cite{gundawar2024robust}.
Those results suggest there is still room to explore how best to harness \glspl*{LLM} in these complex settings which require more advanced reasoning.
While we anticipate that performance will improve as \glspl*{LLM} continue to evolve, we also hypothesize that, like human experts, these systems will demonstrate improved performance when provided access to domain-relevant ``tools''.
Here, ``tool'' is used in a very broad sense and includes (but is not limited to): custom datasets, computational models, the ability to execute a physical experiment, or have interactions with humans or other specialized \glspl*{LLM}.

While our long-term objective is to develop a tool-based agent for lunar mission planning, this paper focuses on one specific tool likely to play a key role in such a system: a structured dataset containing detailed geological and compositional information. 
For lunar materials science, accurate representations of composition is critical as it drives a number of downstream scientific studies, such as regolith simulant development and geologic map development.
Although existing datasets provide remotely sensed data from orbital missions, we aim to augment these with additional information currently available only in scientific literature, such as  detailed mineralogical and trace element analyses derived from in situ sample analysis.
To this end, we explore the task of automatically extracting compositional data from a representative collection of scientific documents using \glspl*{LLM}. 
In particular, we examine the complexities of accurately extracting knowledge in cases where there is variability in composition.
Progress in this area represents a valuable step toward achieving our broader goal of enabling robust lunar mission planning.

\section{Related Work} \label{sec:relatedwork}

One common paradigm for leveraging a document corpus with \glspl*{LLM} is \gls*{RAG}, which refers to the process of dynamically identifying a small subset of relevant information from the corpus and providing this to the \gls*{LLM} within the context window at query time as part of the overall prompt~\cite{gao2023retrieval,fan2024survey}.
\gls*{RAG} thus provides a minimally invasive mechanism to augment the parametric knowledge encoded within the \gls*{LLM} weights.
The most basic implementation of \gls*{RAG} involves two steps. 
First, a preprocessing step partitions documents into small subsets (``chunks'') and embeds these into a space that facilitates efficient comparison (e.g., some vector space informed by the \gls*{LLM}'s latent representation).
The second step then compares these vectors to an embedding of the \gls*{LLM}'s prompt (e.g., via cosine similarity) and then augments the prompt using the top $k$ best matches.
While \gls*{RAG} based on chunking and vector-based retrieval offers flexibility, it can also introduce challenges.
First, partitioning documents arbitrarily runs the risk of separating pieces of information that may jointly be required to answer some query.
Fragmenting a table is a trivial example; a more subtle example is that identifying certain relationships may involve combining information from multiple paragraphs or sections of the paper. 
While more informed chunking schemes are possible, e.g., based on document layout analysis, it is still not trivial to anticipate all the ways relevant information may be distributed across document sections.
A second challenge is that it can be challenging to devise a retrieval step so that the right chunks are prioritized and occupy the top $k$ set of results.

When the downstream queries can be reasonably anticipated and the size of the document corpus is modest, an alternative approach is to identify and extract the core pieces of information that will be relevant to a wide variety of these queries and place them into a structured form (e.g., a relational or graph database) during preprocessing (\cref{fig:architecture}).
While constructing such a structured representation is more time-consuming than simple ``chunking'', structured representations can make it easier for a human to inspect and modify data points (versus, say a vector embedding which is opaque).
Additionally, databases enable queries that can combine multiple pieces of information in nontrivial ways, e.g., the SQL language provides multiple facilities for joining, filtering, and performing arithmetic operations on data.

However, structured representations need not be constructed manually --- a relative strength of \glspl*{LLM} is the ability to work with natural language and interpret linguistic context making them a prime candidate for automating this preprocessing step.
Systematically analyzing each document with an \gls*{LLM} is substantially faster and cheaper than manual human analysis, allowing this approach to scale to relatively large collections of documents.
To date, there have been a number of studies that explore this approach.
Examples of applying \glspl*{LLM} for extracting structured scientific data include~\cite{polak2024flexible,polak2024extracting,schilling2024text,edge2024local}.
There have been a number of works focused upon specific nuances associated with data extraction, e.g., table recovery~\cite{lu2025large}.

\section{Approach} \label{sec:approach}
\begin{figure}
\centering
\includegraphics[width=\textwidth]{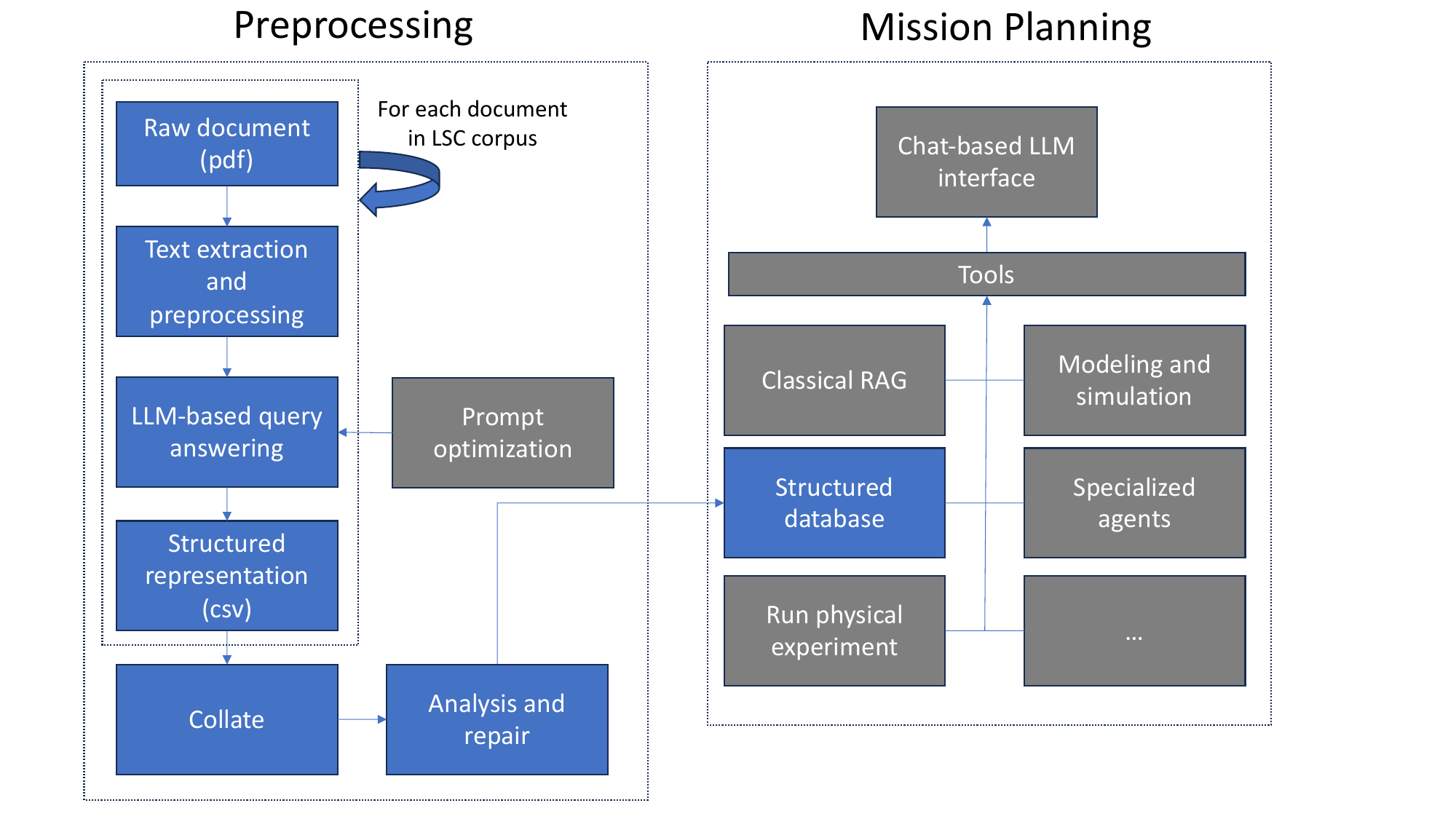}
\caption{Envisioned tool-enabled \gls*{LLM} for mission planning. In this paper, we focus on the preprocessing task of extracting structured information from scientific documents (blue boxes) which can then serve as a source of auxiliary knowledge to support interactive mission planning (future work). The collection of tools listed here is non-exhaustive.}
\label{fig:architecture}
\end{figure}

We take the approach detailed in~\cref{fig:architecture} of using a ``preprocessing'' \gls*{LLM} to systematically extract structured data representations from a corpus of scientific publications.
Specifically we implement a two-step process whereby, for each lunar publication, we first extract all the text from the pdf file using conventional software libraries and then prompt the \gls*{LLM} to summarize the composition details in tabular form (a relatively simple structured representation).
We do not currently attempt to process images or figures.

\subsection{Dataset} \label{sec:dataset}
\begin{figure}
\centering
\includegraphics[width=.75\textwidth]{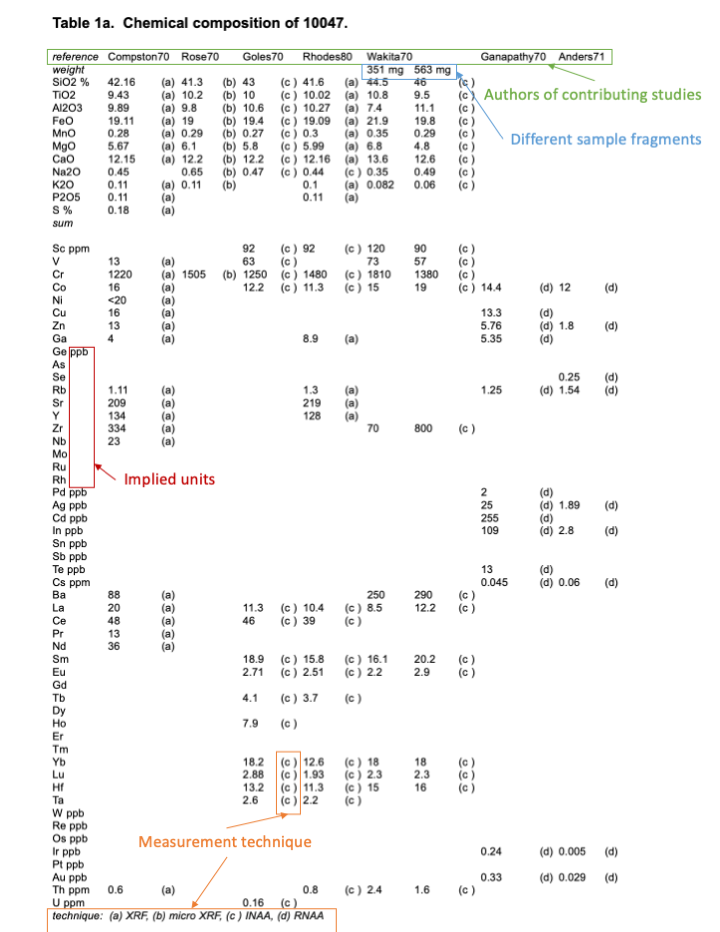}
\caption{
Example of chemical composition data from \gls*{LSC} document \texttt{10047.pdf}. 
Note these tables exhibit various complexities, including analyses by multiple authors, blank entries, multiple units (percent, \gls*{ppm}, \gls*{ppb}), and rows where the units are implied based upon entries above.
}
\label{fig:lsctable10047}
\end{figure}

We take as a representative problem the task of reliably extracting composition information from lunar scientific publications.
For our corpus, we elected to use a subset of the \gls*{LSC}~\cite{meyer2005lunar} corresponding to detailed analyses for lunar samples collected from the Apollo 11, 12, 14, 15, 16, and 17 missions\footnote{The \gls*{LSC} also has documents available for download related the Russian Luna mission, but the Apollo documents represent the bulk of the compendium.}.
We downloaded 728 pdf documents from the \gls*{LSC} website~\cite{LSCweb} and manually annotated chemical composition information for ten samples\footnote{Our ground truth was derived from \gls*{LSC} documents: \texttt{10047.pdf}, \texttt{10062.pdf}, \texttt{12004.pdf}, \texttt{12057.pdf}, \texttt{14321.pdf}, \texttt{15415.pdf}, \texttt{15555.pdf}, \texttt{61016.pdf}, and \texttt{71595.pdf}. Note that \texttt{71595.pdf} includes data for both samples 71595 and 71576.} to use for ground truth.
This chemical composition information is primarily contained within tables in the documents which represent summaries from multiple experiments; one such example is provided in \cref{fig:lsctable10047}. 
Our ground truth contains at least one sample from each of the aforementioned Apollo missions\footnote{
The leading digits of the sample id indicate the associated Apollo mission: ``10'' for Apollo 11, ``12'' for Apollo 12, ``14'' for Apollo 14, ``15'' for Apollo 15, ``6'' for Apollo 16, and ``7'' for Apollo 17.
}.

The chemical composition tables within these documents typically include information from multiple studies. In addition to variability arising from multiple research groups, the samples themselves are not homogeneous but polymineralic rock; i.e., they are composed of a variety of minerals, often with different physical characteristics, all intergrown together to form the rock structure.
Consequently, we elect to represent composition for a lunar sample as an interval quantity defined as the range between the minimum and maximum reported value for each oxide or element.
Future work might explicitly represent these via a probabilistic distribution of phases and/or attempt to explicitly characterize each mineralogical phase individually.

\begin{figure}
\centering
\includegraphics[width=.75\textwidth]{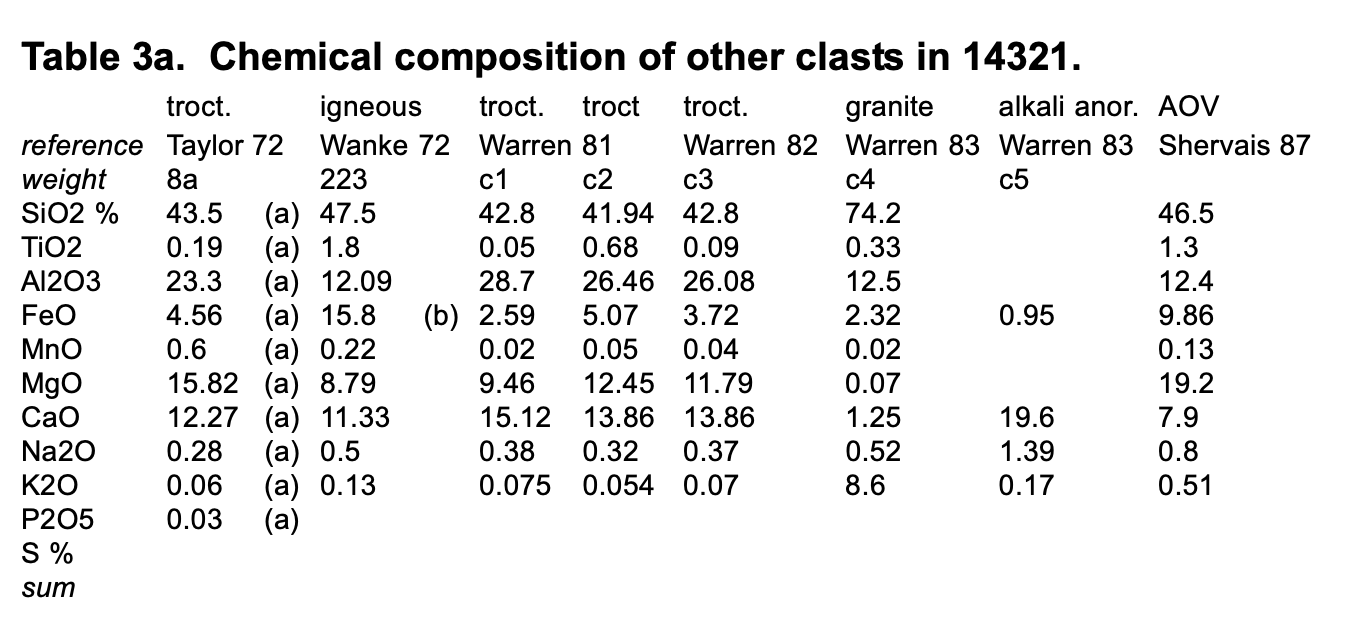}
\caption{Excerpt from the \gls*{LSC} document for sample 14321.  Here, the columns denoting different studies also designate the different mineralogical phases (termed ``clasts'' in the column header) derived from this sample. Our current ground truthing and \gls*{LLM} prompting strategy does not attempt to disambigute among the various phases within a sample; however, this would be a natural next step and has implications for performance analysis. For example, the ``granite'' column demonstrates significant compositional differences relative to the others.
}
\label{fig:LSC_14321_table3a}
\end{figure}

\subsection{Implementation Details}
To implement the preprocessing pipeline we first used the \texttt{PyMuPDF}~\cite{pymupdf2025} Python package to extract text from each sample document.
We processed each document independently, providing the full content of the paper to the \gls*{LLM} using the prompt  shown in~\cref{lst:prompt}.
Note that in our prompt the \gls*{LLM} is explicitly asked to summarize values across multiple measurements and provide the answer as an interval quantity.
We originally experimented with employing simple document layout analysis strategies during this preprocessing step (e.g., to omit the references portion of the document and to explicitly annotate various portions of the document as tables, figure captions, plain text, etc.) but found that \ChatGPT performed well without the aid of this additional preprocessing.
The results (presented later, in~\cref{sec:results}) are of course a function of the prompt and, while  we experimented empirically with different versions of this prompt, there is likely opportunity for further refinement and improvement (possibly leveraging optimization tools, e.g.,~\cite{khattab2023dspy}).

Due to token rate limitations, we were not always able to provide the entire document body within a single context window. 
In these cases we: split the document into pieces of approximately 25,000 characters while respecting page boundaries, ran the query on each ``sub-document'', and then aggregated the \gls*{LLM} results as a post-processing step.
While splitting documents has the potential to be deleterious (as per our discussion on chunking), we anecdotally observed cases where this was beneficial. 
For example,  in the case of sample 14321, we observed improved end results; this document contained multiple tables describing composition, and a hypothesis is that providing these to the \gls*{LLM} in separate invocations simplified the extraction process. 
There has also been work on how the relative location of information within the context window impacts performance (e.g.,~\cite{liu2024lost}) so it is also possible that this was a factor.
A more systematic investigation of preprocessing strategies is an opportunity for future work.

For our experiments we used \ChatGPT with the temperature parameter set to 0. 
Prior to running our analyses, we implemented a few minor data cleaning and repair steps to the raw output from the \gls*{LLM}. 
This included deduplicating entries when the \gls*{LLM} produced multiple values for a given (sample, composition) tuple, light reformatting (e.g., in cases where the \gls*{LLM} reported a single value instead of an interval as instructed), and removing non-numeric content from sample ids (e.g., in some cases the \gls*{LLM} reported both the sample number and the the associated mineralogical phase as the sample identifiers).
In all cases, we kept the composition values faithful to what was reported by the \gls*{LLM}.

\begin{figure}
\begin{lstlisting}[
frame=single,
caption=Prompt used in our experiments.,
label=lst:prompt,
breaklines=true,
breakautoindent=false,
breakindent=0ex,  
basicstyle=\ttfamily \footnotesize,
]
The document text below contains data related to one or more Apollo lunar samples. Please extract all the chemical composition information. If there is no obvious composition content in text, it is acceptable to return an empty table.

Specifically, list the abundance percentages (%) of the compounds and/or standalone elements that are present in the sample. Note that a given sample may have been analyzed by multiple groups, so there may not be a single value but an entire range of values for each element or compound. You must format your response as a comma separated value table, with one row for each element or compound. Columns should consist of the element or compound in question, the corresponding Apollo sample id, the weight fraction as a range of observed values, and finally a column for the units. Typical units will be percent (%), parts per million (ppm), or parts per billion (ppb). If you would like to list a single percentage instead of a range, write it as XX-XX%, for example 55-55%.

It is okay if you can only give your best guess at composition percentages, if you are not sure on the exact answer. Either way, you MUST answer. Do not respond with any other text besides this table. An example of a valid response might be:

Compound, SampleId, weight, units
SiO2, 15535, 44.46-45.5, percent,
TiO2, 15535, 2.15-2.51,  percent,
Cr,   15535, 3900-5094,  ppm,
SiO2, 15536, 44.1-44.6,  percent,
TiO2, 15536, 2.14-2.7,   percent,
Cr,   15536, 4100-6419,  ppm
\end{lstlisting}
\end{figure}

\subsection{Evaluation Metrics} \label{sec:metrics}
As demonstrated in \cref{fig:LSC_14321_table3a}, lunar samples are not homogeneous.
Furthermore, both materials formation and characterization processes typically have associated variability.
Therefore, ``ground truth'' in this setting is better represented via interval values as opposed to single points
and our metrics will include assessments of these interval quantities.
We focus here solely on the performance of this extraction step and defer to future work analysis of \gls*{LLM} queries that involve drawing inferences across multiple lunar samples.
In this section we define metrics for comparing two interval quantities.

For closed intervals of the reals $A = [a_0, a_1], a_1 > a_0$ we denote the length of A via
\begin{equation}
|A| = a_1 - a_0,
\end{equation}
and the midpoint of $A$ by
\begin{equation}
m_A = a_0 + (a_1 - a_0) / 2.
\end{equation}
Let $T$ denote the ground truth interval and $E$ a corresponding estimate. Two ways to compare their ``central tendency'' are the absolute difference in midpoints
\begin{equation} \label{eq:mp_abs_delta}
\delta_{E} = |m_T - m_E|,
\end{equation}
and the relative percent difference in midpoints
\begin{equation} \label{eq:mp_abs_delta_rel}
\delta_{E}^r = 100 \cdot \frac{|m_T - m_E|}{ m_T}.
\end{equation}
To account for the extent of intervals, we also measure performance using precision and recall
\begin{equation} \label{eq:precision}
\text{precision} = \frac{\left|T \bigcap E \right|}{|E|},
\end{equation}
\begin{equation} \label{eq:recall}
\text{recall} = \frac{\left| T \bigcap E \right|}{|T|}.
\end{equation}
We compute these metrics only in settings where the ground truth interval $T$ exists.
There can also occur ``false positives'' when the \gls*{LLM} reports values for compositions that are not included as part of the ground truth.
Therefore, we also present in the Appendix supplementary analyses that consider these scenarios.

\section{Results} \label{sec:results}

In the following we compare two sets of \gls*{LLM}-derived results to ground truth.
The first evaluates the performance of the ``preprocessing'' pipeline in~ \cref{fig:architecture}, where the entire corpus is presented, one paper at a time, to a \gls*{LLM} and the collated results are then evaluated (i.e., within the ``analysis and repair'' block).
This approach is termed ``with doc.'' in forthcoming plot legends to denote that the \gls*{LLM} was provided the corpus content directly. 
For comparison, we also include a baseline where an independent \gls*{LLM} is prompted for results without providing any content from the \gls*{LSC} corpus (termed ``standalone'' in the legends).
This comparative baseline is included to assess whether the data of interest might already be captured directly by the \gls*{LLM}'s internal parametric representation.\footnote{Since these \gls*{LSC} documents are in the public domain, it is reasonable to assume they may have been included as part of any \gls*{LLM}'s training corpus.}

Given the aforementioned nuances of the tables appearing in the lunar documents, and the fact that data extraction is not just a simple lookup but also involves some aggregation of numerical information from multiple measurements, it was not clear \textit{a priori} how successful either approach would be.
Overall we find that the \gls*{LLM}, when provided with the paper, achieves less that 5\% relative error for the majority of the points we ground truthed, and that this is significantly better than querying the \gls*{LLM} in isolation (i.e., without providing the paper).
The remainder of this section details specific results.

\subsection{Qualitative Analyses}
\begin{figure*}[t!]
    \centering
    \includegraphics[width=\linewidth]{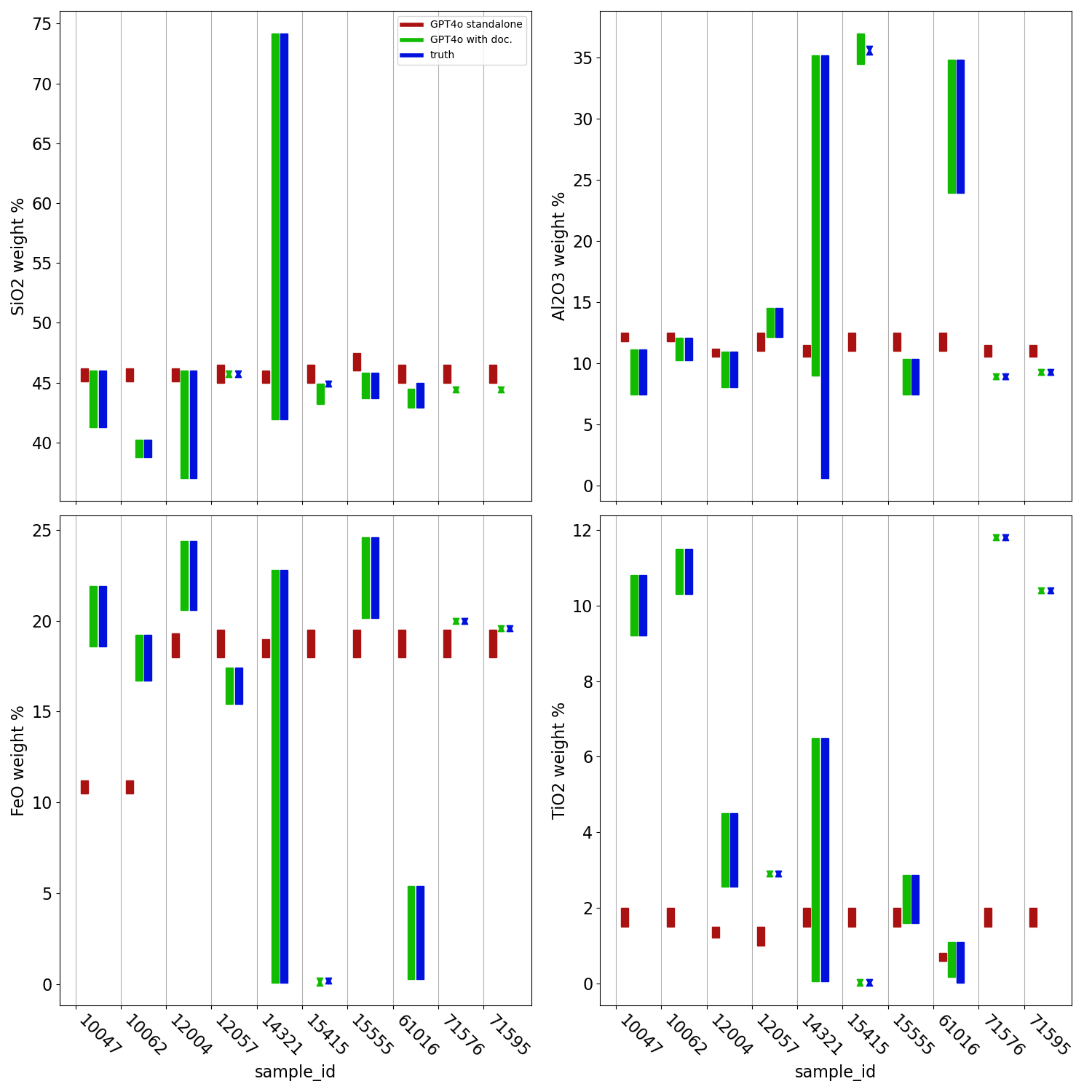}
    \caption{
    Weight percentages for four relatively abundant oxides. Blue intervals denote the manually extracted ground truth, green values denote data extracted by \ChatGPT when the paper content is included within its context window, and red denotes the baseline \ChatGPT result when the paper content is not provided. The green and blue intervals demonstrate much closer alignment across the entire range of samples.
    When an interval is small relative to the y-axis, it is rendered as a an hourglass-shaped marker rectangle (e.g., the blue and green values \ch{SiO2} for sample 12057).
    Note the x-axis is the same for all subplots whereas the y-axis is not.
    Also note the lack of explicitly defined ground truth for \ch{SiO2} in samples 71576 and 71595; in these cases, the \glspl*{LLM} have provided an answer (despite the table entry being blank, see~\cref{fig:lsctable71595}). 
    }
    \label{fig:per-material-intervals}
\end{figure*}

In order to contextualize the performance metrics, it is beneficial to first visualize the intervals themselves. \Cref{fig:per-material-intervals} depicts the results for four oxides that are relatively abundant in the lunar samples for which we have ground truth.
In general, \gls*{LLM} outputs when equipped with the paper contents (green intervals) visually demonstrate better agreement with the ground truth (blue intervals) relative to the \gls*{LLM} in isolation (red intervals). 
Note also the ``standalone'' results tend to exhibit a relative lack of sensitivity to sample id, e.g., for \ch{SiO2} and \ch{Al2O3}, all of the red intervals appear in the vicinity of 46\% and 12\% respectively. 

\begin{figure*}[t!]
\centering
\includegraphics[width=.9\linewidth]{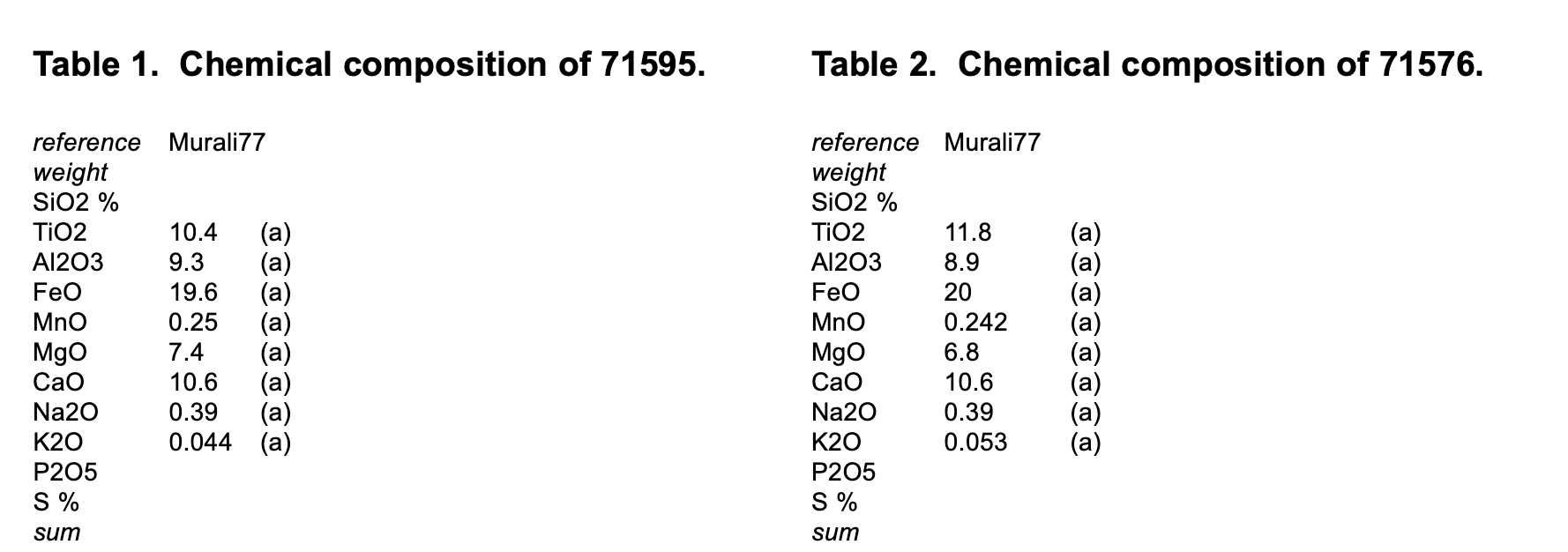}
\caption{
Excerpt from \gls*{LSC} document \texttt{71595.pdf}. Note that \ch{SiO2} quantities are not explicitly reported for these samples (and, hence, not included in ground truth; see also~\cref{fig:per-material-intervals}).
}
\label{fig:lsctable71595}
\end{figure*}

\subsection{Quantitative Performance}

\begin{figure*}[ht]
    \centering
    \begin{subfigure}[t]{0.49\textwidth}
        \centering
         \includegraphics[width=\linewidth]{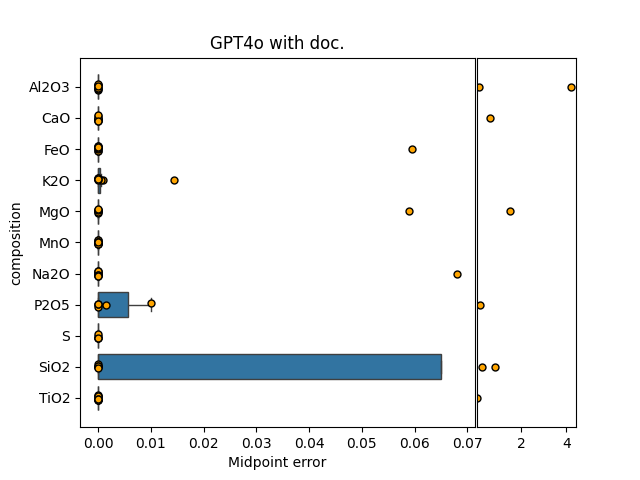}
        \caption{Performance using paper content.}
    \end{subfigure}%
    ~ 
    \begin{subfigure}[t]{0.49\textwidth}
        \centering
         \includegraphics[width=\linewidth]{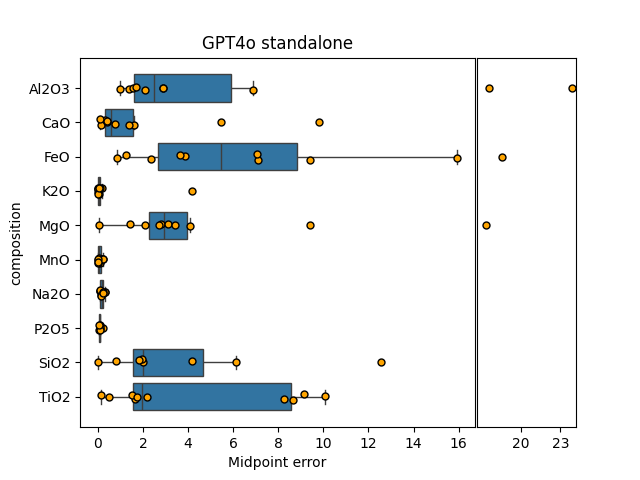}
        \caption{Standalone performance.}
    \end{subfigure}
    \caption{\gls*{LLM} data extraction performance for non-trace compositions. 
    The left panel shows performance when the \gls*{LLM} is provided with the \gls*{LSC} content while the right panel shows performance when the \gls*{LLM} is queried in isolation. The performance metric is absolute midpoint error, as defined in~\cref{eq:mp_abs_delta}.
    The box plots show the distribution of scores, where there is a score for each (sample, composition) pair.
    Individual scores are overlaid on the box plots as orange dots and the $x$-axis is ``broken'' to keep large outliers from compressing the range. 
    Note the different scale of the $x$ axes, highlighting that providing the documents leads to improved performance.
    }
    \label{fig:perf-abundance-midpoint}
\end{figure*}

For our first quantitative analysis, we consider the level of agreement between the midpoints of true vs.\ \gls*{LLM} extracted intervals. 
\Cref{fig:perf-abundance-midpoint} shows the results for all non-trace compositions.\footnote{In this context, we define ``non-trace'' compositions as those whose weights are reported in units of percentages (vs. \gls*{ppm} or \gls*{ppb}) within the source \gls*{LSC} documents. With the exception of \ch{S}, these are all oxides.}
We observe that the inliers exhibit an absolute error of less than 1\% and a relative error of less than 5\%.
The figure also demonstrates the utility of providing the paper to the \gls*{LLM} within its context window as performance is superior to the \gls*{LLM} in isolation. 
This is consistent with intuition that---when a specific result associated with a particular corpus is sought---providing the \gls*{LLM} with content from that corpus is likely to lead to improved performance.
These results also mirror what we observe in~\cref{fig:per-material-intervals}.

\begin{figure*}[t!]
    \centering
    \begin{subfigure}[t]{0.49\textwidth}
        \centering
         \includegraphics[width=\textwidth]{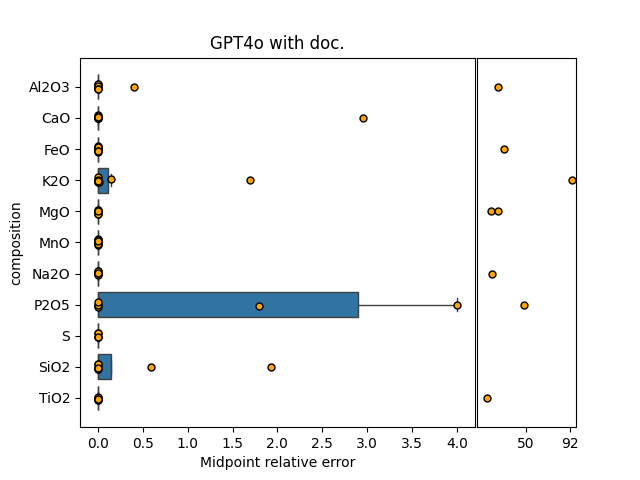}
        \caption{Performance using paper content.}
    \end{subfigure}%
    ~ 
    \begin{subfigure}[t]{0.49\textwidth}
        \centering
         \includegraphics[width=\textwidth]{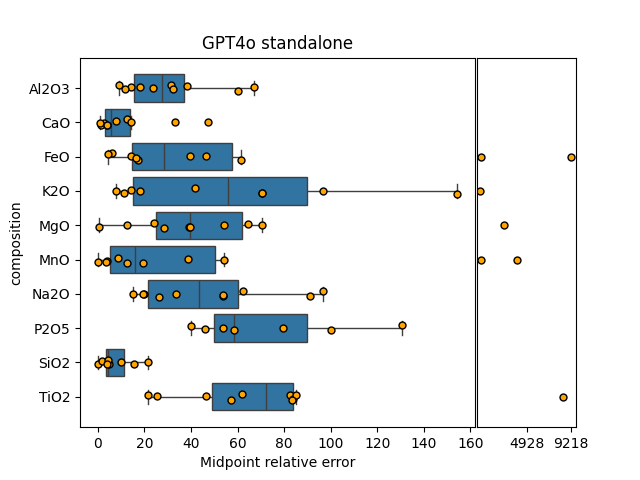}
        \caption{Standalone performance.}
    \end{subfigure}
    \caption{
    \gls*{LLM} data extraction performance measured in percent relative error, i.e.,~\cref{eq:mp_abs_delta_rel}. 
    As in \cref{fig:perf-abundance-midpoint}, providing the document leads to much better agreement with the ground truth, however now we observe errors associated with \ch{P2O5} contribute more in a relative sense vs absolute.
    }
    \label{fig:perf-abundance-midpoint-relative}
\end{figure*}

Since there are substantial differences in abundances among compositions (e.g., \ch{SiO2} tends to be more abundant than other oxides), it is also useful to consider the relative percent error as in~\cref{eq:mp_abs_delta_rel}, depicted in \cref{fig:perf-abundance-midpoint-relative}.
When provided with the paper content, the relative error tends to be below 5\%, although there are a number of significant outliers.
However, we note that this metric is sensitive to cases where the denominator of~\cref{eq:mp_abs_delta_rel} is small.
For example, the \ch{FeO} relative error outlier of $\approx 30$\% in the left panel of~\cref{fig:perf-abundance-midpoint-relative} corresponds to sample 15415 where the ground truth interval is $[0.199, 0.202]$ and the \gls*{LLM} estimate was $[0.08, .202]$ (see also~\cref{fig:per-material-intervals}).
The estimated interval contained the ground truth, but incurred precision error which in turn substantially impacted the midpoint relative error.
Thus, relative percent error does not always convey the whole story when it comes to interval performance.

\begin{figure*}[ht]
    \centering
    \begin{subfigure}[t]{0.49\textwidth}
        \centering
         \includegraphics[width=\linewidth]{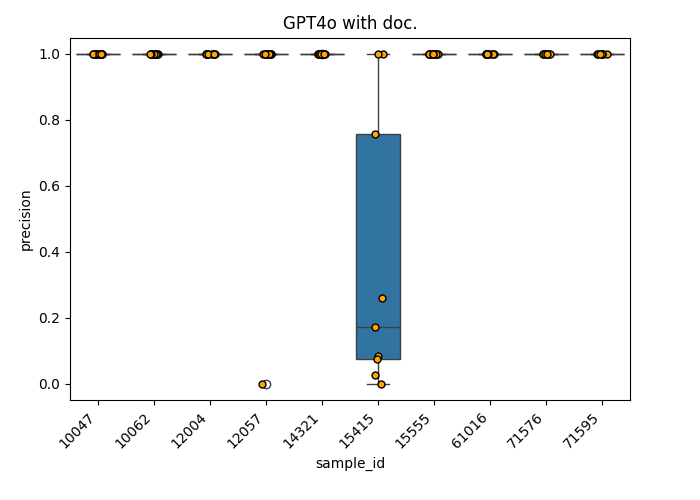}
        \caption{
        Precision scores per sample id.
        }
    \end{subfigure}%
    ~ 
    \begin{subfigure}[t]{0.49\textwidth}
        \centering
         \includegraphics[width=\linewidth]{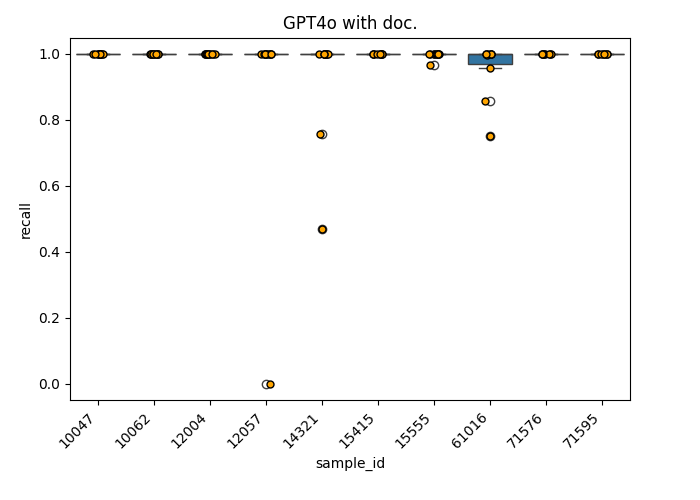}
        \caption{
        Recall scores per sample id.
        }
    \end{subfigure}
    \caption{
    Precision and recall performance (when provided with \gls*{LSC} papers) for most abundant compositions. Individual scores (i.e., for each oxide) are shown as orange dots while the overall distribution of scores is characterized by the box plot.
    The behavior of sample 15415 is considered further in~\cref{fig:15415}.
    }
    \label{fig:perf-precision-recall}
\end{figure*}

This motivates~\cref{fig:perf-precision-recall} where we analyze interval agreement through the lenses of precision and recall.
There is very good agreement across most of the samples; most of the precision errors were associated with sample 15415 and recall performance was generally good, although we do observe a few substantial outliers.

\begin{figure*}[ht]
    \centering
    \begin{subfigure}[t]{0.5\textwidth}
        \centering
        \includegraphics[width=\textwidth]{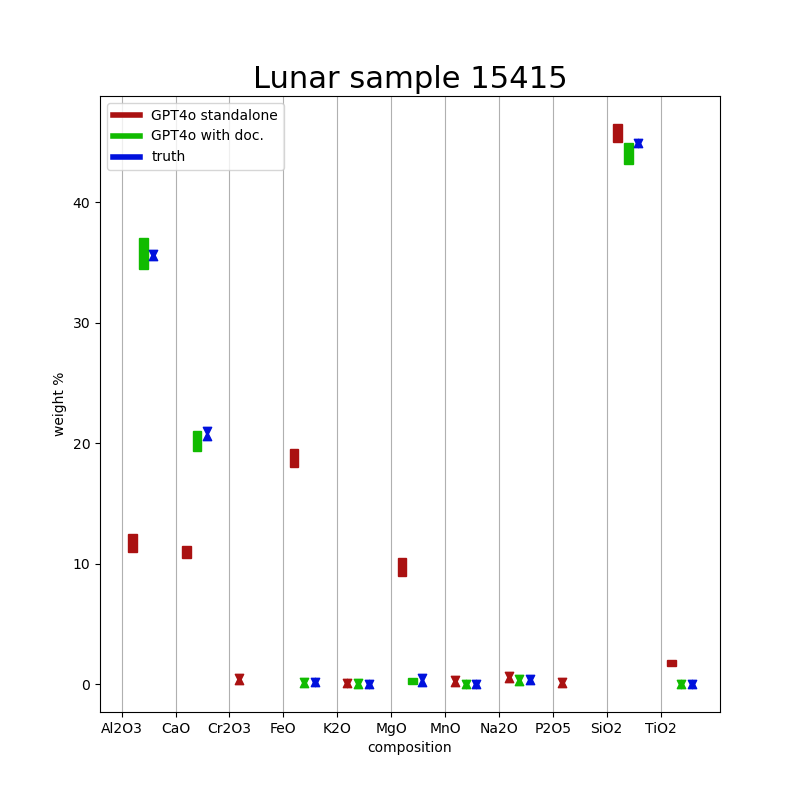}
        \caption{Composition intervals.}
    \end{subfigure}%
    \begin{subfigure}[t]{0.5\textwidth}
        \centering
        \raisebox{0.7cm}{
            \includegraphics[width=\textwidth]{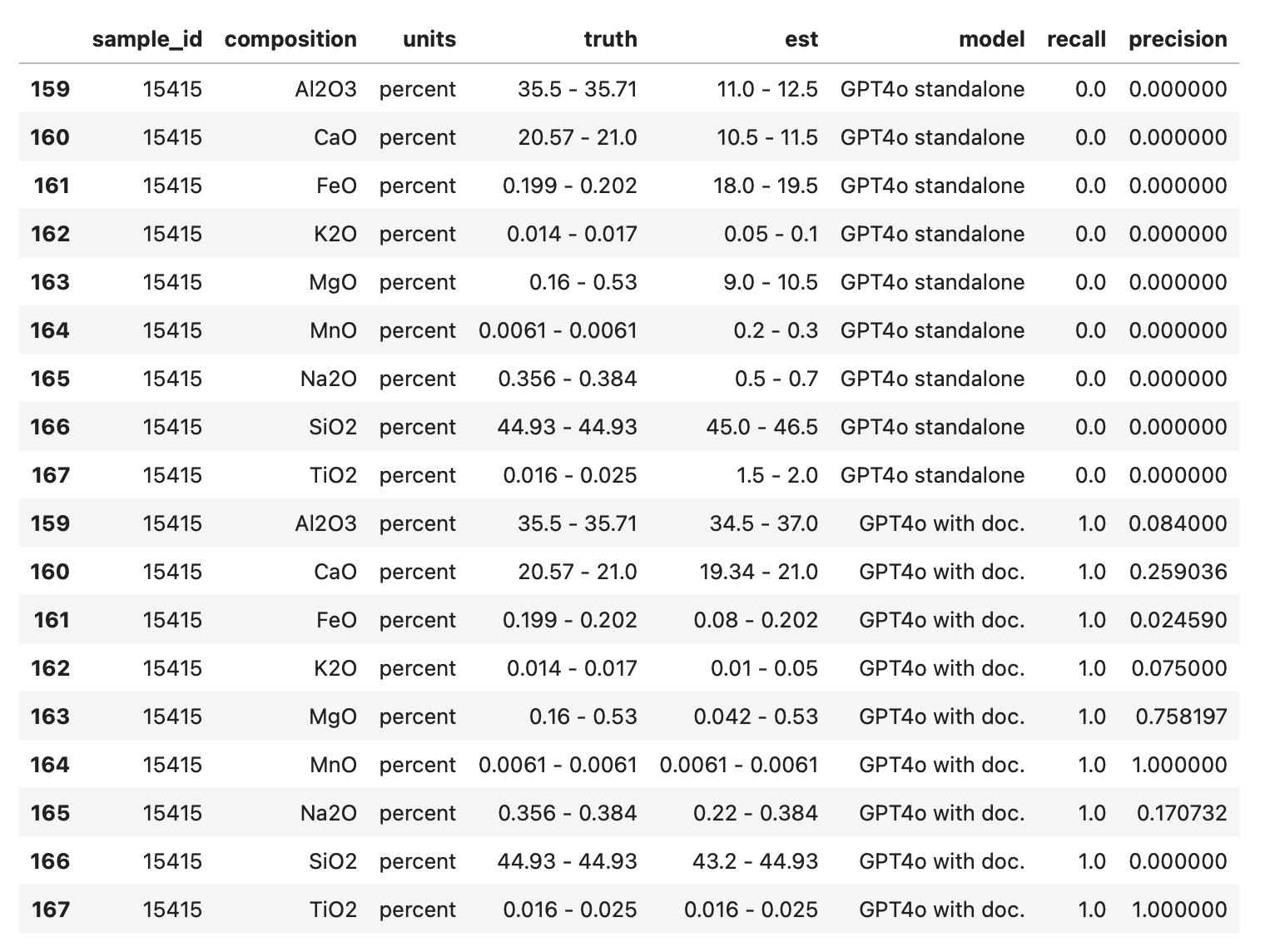}
        }
        \caption{Underlying truth and LLM estimates for sample 15415 (used to generate panel a).}
    \end{subfigure}%
    \caption{
    Estimated compositions vs.\ ground truth for sample 15415, where the \gls*{LLM} ``with document'' precision was weakest. Note in the left panel that \ch{P2O5} does appear in table 1 \texttt{15415.pdf} but without any reported value and \ch{Cr2O3} is not listed as an abundant oxide in that same table. Thus, the ``standalone'' \gls*{LLM} values reported for these might be considered ``false positives'' from the perspective of our ground truth (even if they are present in reality).
    }
    \label{fig:15415}
\end{figure*}

In the case of sample 15415, we observe in~\cref{fig:15415} that the ground truth values for many of these compositions are tightly concentrated.
In some cases, such as \ch{Al2O3}, \ch{CaO}, and \ch{SiO2}, the \gls*{LLM} with paper (green intervals) is incurring high precision errors as the predicted interval is substantially larger than the ground truth.
In other cases, such as \ch{FeO} and \ch{K2O}, both the extracted and true intervals are in fairly good agreement (again, for the \gls*{LLM} with paper), but the relative nature of the calculation results in a significant precision penalty.
The \gls*{LLM} without the paper (red intervals) provides responses that differ substantially from the ground truth.
It is likely that other domain-inspired metrics that take into account the true tolerance for error for a given application (e.g., specific to regolith simulant development or a particular desired synthesis process) might also be developed to provide other perspectives on performance.

The evaluations in this section focused upon the most abundant compositions, and are largely focused on comparing \gls*{LLM} estimates to the small subset of examples which we manually truthed. In~\Cref{sec:aux-results}
we provide additional analyses across the set of 700+ documents that we downloaded from the \gls*{LSC} corpus.

\section{Conclusions and Future Work} \label{sec:conclusion}

We have evaluated the capabilities of \ChatGPT in extracting relevant materials science data from lunar scientific publications, with the broader goal of supporting a mission planning agent.
Our initial findings are largely encouraging, though several potential challenges stem from the complexity of the underlying data (e.g., multiple mineral phases) or the sensitivity of interval metrics (e.g., when ground truth is more point-like).
We hypothesize that as these models continue to advance, their performance in this domain will improve. 
A natural next step is to evaluate performance using next-generation \glspl*{LLM} beyond \ChatGPT.

In this work we performed minimal preprocessing to the documents before providing them to the \gls*{LLM} for data extraction. 
We mentioned above early versions of this work where we also experimented with document layout analysis as part of document preprocessing.
While this did not substantially improve performance, 
we did anecdotally observe instances where a very coarse partitioning of the document provided some a benefit. 
This suggests there is opportunity to revisit document preprocessing, although the impact may depend upon the structure of the paper and nature of the information the \gls*{LLM} is being asked to extract. More controlled and rigorous studies into this, especially as it relates to the \gls*{LLM} context window, could be fruitful. 
This would include looking at other datasets, beyond the \gls*{LSC}.

Another key direction is assessing how well \glspl*{LLM} can extract more fine-grained results, such as mineralogy-specific composition data (recall~\cref{fig:LSC_14321_table3a}). 
Expanding ground truthing to include data embedded in the body of the text, and in less structured form than tables, would further enrich our understanding of model performance. Ultimately, a critical question is whether these models can synthesize conclusions from multiple pieces of information spread across one or more documents.
Simple examples might include queries such as ``what is the average iron content near South Ray Crater?'' (requires sample localization and simple averaging); more sophisticated queries could entail fusing composition information with geological models.

Future work could explore how recent advances in prompt engineering and tool-enabled agent optimization improve both data extraction and \gls*{LLM}’s ability to integrate this data into mission planning workflows.
Given that entirely error-free extraction is unlikely—especially for queries requiring complex ``multi-hop'' reasoning—it is also worth investigating how \gls*{LLM} uncertainty measures, combined with adaptive sampling strategies, could help direct human annotators to review the most uncertain pieces of extracted information.

Integration into our agent-based system remains in its early stages. However, ongoing development of this tool and its associated data representations has the potential to transform how we collect, process, and utilize lunar materials. This rapid, on-demand tool could play a crucial role in enabling sustainable lunar missions, optimizing resource utilization, and fostering a thriving lunar economy.

\section{Acknowledgements}
The authors would like thank the members of the McQueen Lab's Quantum Materials Research group, most notably Gregory Bassen, Wyatt Bunstine, Noah Edmiston, and Tyrel McQueen, for helpful discussions that positively influenced the direction of this work. We also thank Alexander New and Nam Le from the Johns Hopkins Applied Physics Laboratory for their feedback and gratefully acknowledge internal financial support from the Johns Hopkins Applied Physics Laboratory's Independent Research \& Development Program.

\clearpage
\bibliographystyle{plain}
\bibliography{refs} 

\clearpage
\appendix

\section{Additional results} \label{sec:aux-results}

\subsection{Full Corpus Analysis}

\begin{figure}[ht]
\centering
\includegraphics[width=.9\textwidth]{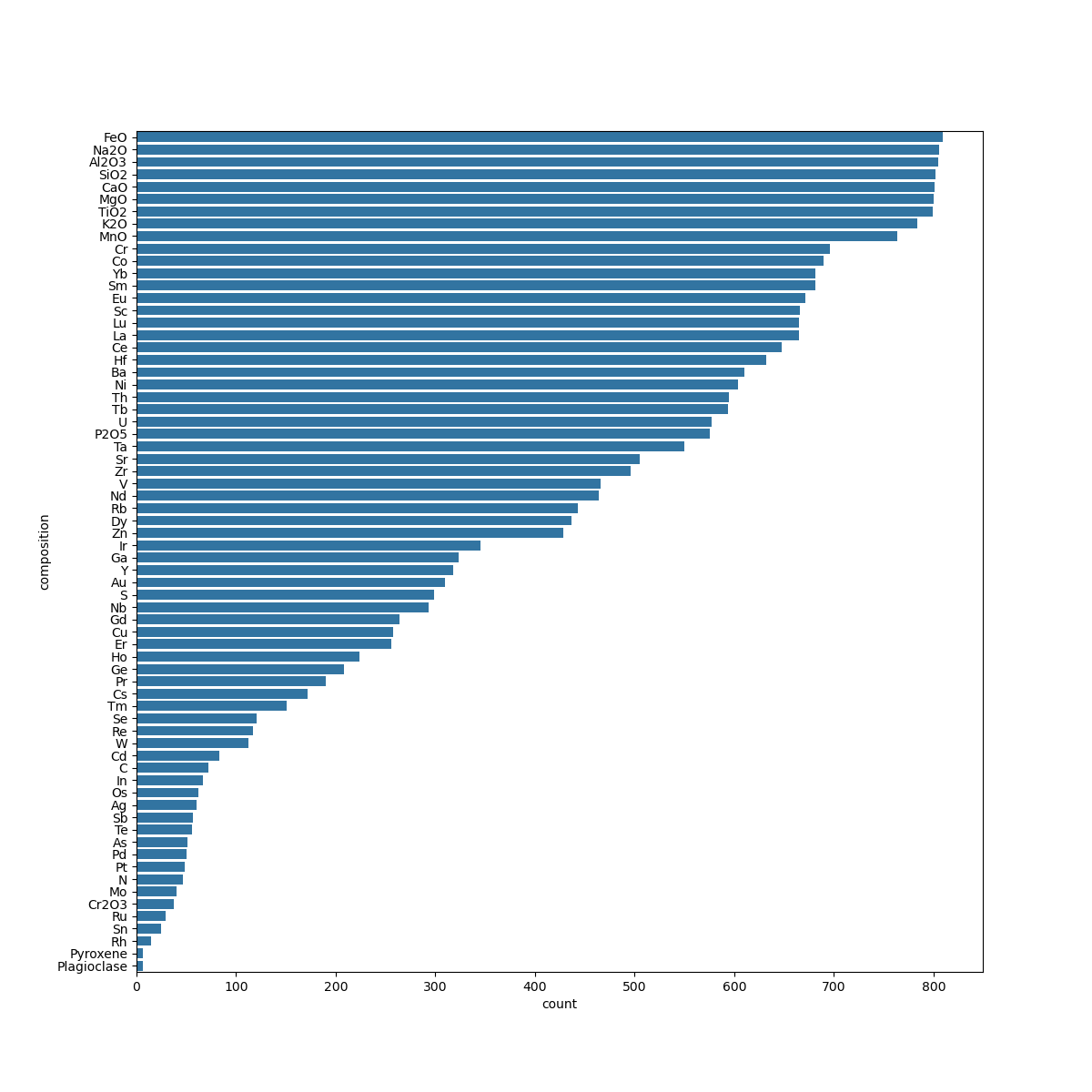}
\caption{
Compounds / elements (y-axis) and number of times an associated interval was extracted by the \gls*{LLM} (x-axis) when applied to the entire subset of the \gls*{LSC} corpus we downloaded.
}
\label{fig:full-corpus-counts}
\end{figure}

In this section we present some qualitative analyses describing what the \gls*{LLM} was able to extract from the 700+ document corpus we used in this study.
\Cref{fig:full-corpus-counts} depicts the frequency and type of composition information extracted.
The figure omits any ``composition'' that appeared five or fewer times in the extracted data; these were presumed to be data extraction errors (not verified). There were 125 such discarded ``compositions''.
It is clear that some mineralogical information was being reported as a composition by the \gls*{LLM} (e.g., see ``Pyroxene'' and ``Plagioclase'' in the y axis of \cref{fig:full-corpus-counts}). The degree to which this can be disentangled via more specific prompting is an area for future work. More broadly, there are doubtless other extraction artifacts and further analysis/work might be done to further improve the results; however it is encouraging that the most frequent oxides appear to be consistent with what is expected.

\begin{figure*}[ht]
    \centering
    \begin{subfigure}[t]{0.49\textwidth}
        \centering
         \includegraphics[width=\linewidth]{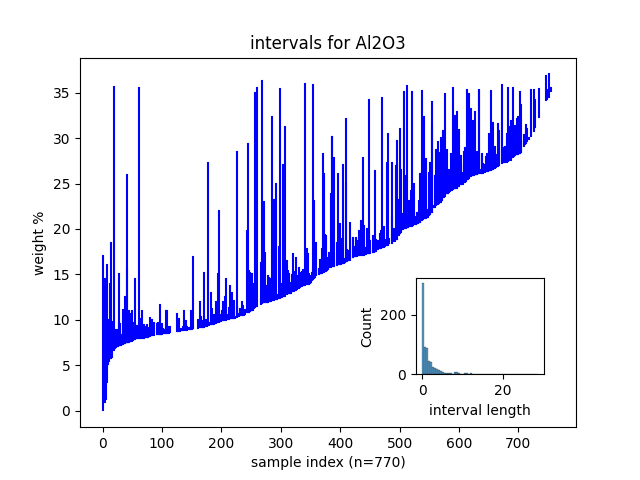}
        \caption{\ch{Al2O3}}
    \end{subfigure}%
    ~
    \begin{subfigure}[t]{0.49\textwidth}
        \centering
         \includegraphics[width=\linewidth]{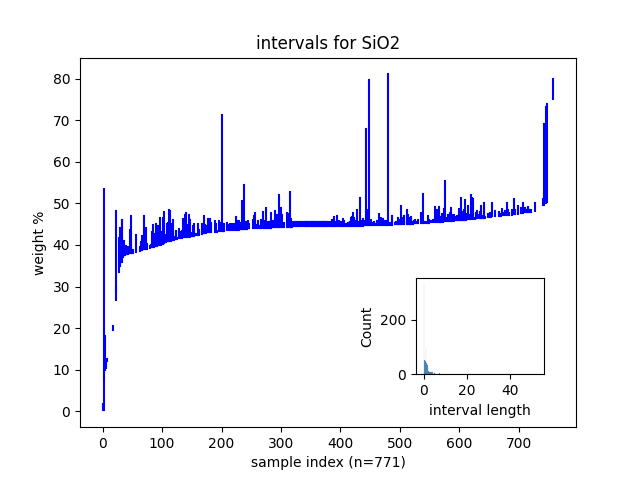}
        \caption{\ch{SiO2}}
    \end{subfigure}
    \vspace{1cm}
    \begin{subfigure}[t]{0.49\textwidth}
        \centering
         \includegraphics[width=\linewidth]{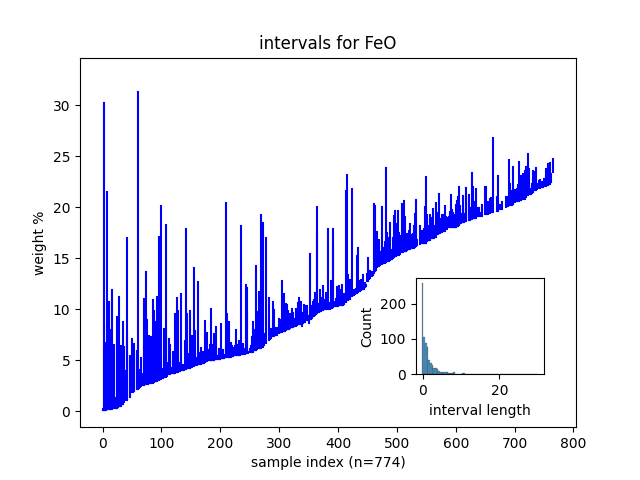}
        \caption{\ch{FeO}}
    \end{subfigure}%
    ~
    \begin{subfigure}[t]{0.49\textwidth}
        \centering
         \includegraphics[width=\linewidth]{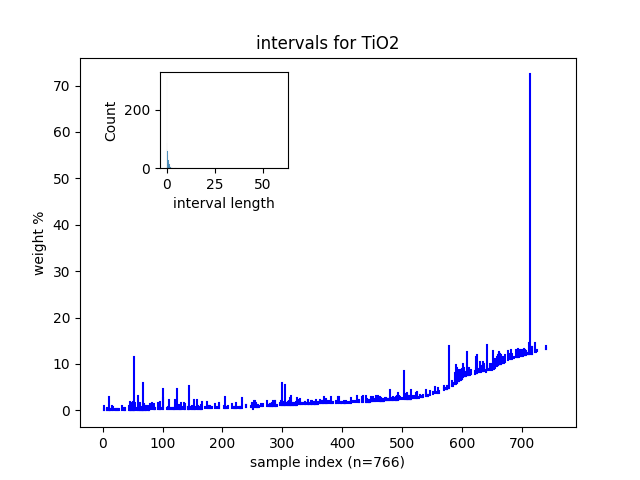}
        \caption{\ch{TiO2}}
    \end{subfigure}
    
    \caption{
    Example weight percentage intervals extracted from the \gls*{LSC} corpus considered in this study. Intervals are ordered along the x-axis by their minimum value.
    Inset plots show the overall distribution of interval lengths. 
    }
    \label{fig:full-corpus-per-material-intervals}
\end{figure*}

We can also consider the range of values (interval widths) that were extracted on a per-compound basis.
\Cref{fig:full-corpus-per-material-intervals} visualizes, for four selected compounds, all of the extracted intervals.
For \ch{SiO2} and \ch{TiO2} for the majority of the documents the weight percentage is fairly range bound (around 45\% and sub-5\% respectively). For \ch{Al2O3} and \ch{FeO} the intervals exhibit much more variability, both in terms of the lower bound and range of the interval.
In future, distributional analyses could provide a useful mechanism for  efficiently cuing human experts towards the pieces of information that would benefit most from manual inspection.

\subsection{Other Analyses}

\begin{figure} 
\centering
\includegraphics[width=.8\linewidth]{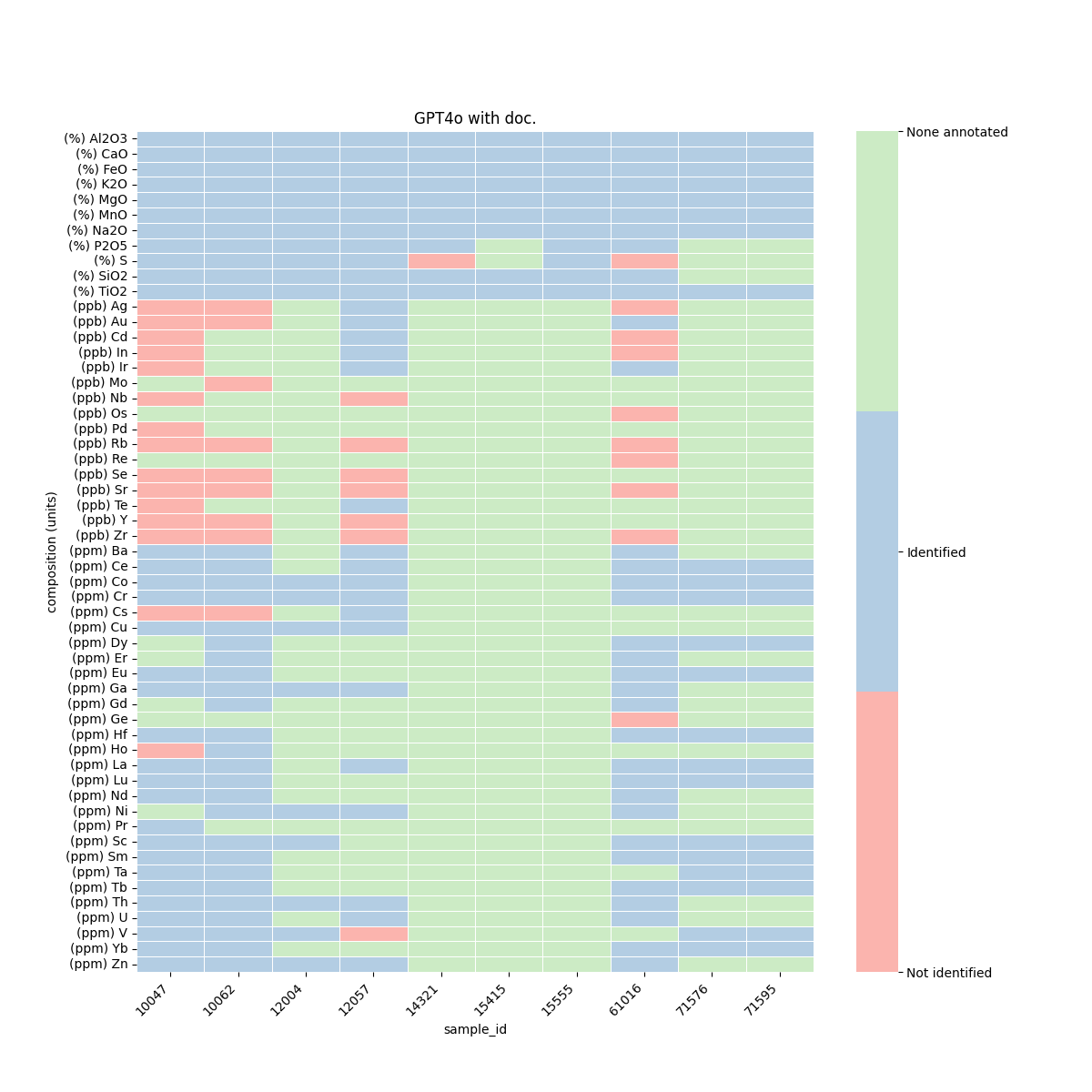}
\caption{Recall for samples that were ground truthed.
Cells shaded in blue denote composition/sample id pairs where the \gls*{LLM} correctly provided some associated interval (independent of accuracy of that interval) while red denotes no corresponding interval was provided despite there being an entry in the ground truth. 
Green cells indicate composition/sample id pairs that were not part of our ground truth. Note that our ground truth was not exhaustive for compounds reported in either \gls*{ppm} or \gls*{ppb}.
}
\label{fig:comp-recall}
\end{figure}

\Cref{fig:comp-recall} depicts how often the ``with document'' \gls*{LLM} returned any interval (regardless of numerical accuracy) for each of the compositions we truthed.
For example, we observe that the \gls*{LLM} missed \ch{S} for samples 14321 and 61016.
The \gls*{ppb} compositions appear to be missed most often; note that, if the \gls*{LLM} incorrectly identifies the units, this is also treated as a miss.
Note also that our ground truthing for \gls*{ppb} and \gls*{ppm} was not exhaustive (whereas for percent we included all composition values from the tables in our truth).

\end{document}